# Does the PBR Theorem Rule out a Statistical Understanding of QM?

Anthony Rizzi
*Institute for Advanced Physics, Baton Rouge, LA 70895*

**Abstract:** The PBR theorem gives insight into how quantum mechanics describes a physical system. This paper explores PBRs' general result and shows that it does not disallow the ensemble interpretation of quantum mechanics and maintains, as it must, the fundamentally statistical character of quantum mechanics (QM). This is illustrated by drawing an analogy with an ideal gas. An ensemble interpretation of the Schrödinger cat experiment that does not violate the PBR conclusion is also given. The ramifications, limits, and weaknesses of the PBR assumptions, especially in light of lessons learned from Bell's theorem, are elucidated. It is shown that, if valid, PBRs' conclusion specifies what type of ensemble interpretations are possible. The PBR conclusion would require a more direct correspondence between the quantum state (e.g., $|\psi\rangle$) and the reality it describes than might otherwise be expected. A simple terminology is introduced to clarify this greater correspondence.

## Introduction

The ensemble interpretation of quantum mechanics, defended by, among others, Albert Einstein, can easily appear to be under threat from a recent discovery.[1] This new discovery is a theorem proved by Pusey, Barrett and Rudolph (PBR).[2] Though later changed, the initial (arXiv) article was titled "*The quantum state cannot be interpreted statistically*."[3] Anthony Valentini says that this theorem shakes the foundations[4] of quantum mechanics. However, the PBR theorem really does not touch the ensemble interpretation. If someone says PBR rules out a statistical interpretation, it is evident that he must also rule out the ensemble interpretation, which is, after all, statistical; in fact, it is prototypical of what one means by statistical. This article will fill out the theorem's real meaning by making it clear that it does not in any way refute the ensemble interpretation. Before moving to our goal of properly understanding PBR by properly understanding its relation to the ensemble interpretation, let me state that PBR is an important result. Nothing said here should be construed as denying or minimizing this fact. Let me also say that PBR, despite earlier versions which we will discuss, does not currently claim that the ensemble interpretation is disproven. However, there is confusion on the issue and they never state that it is not disproven; I hope to clear up the confusion.

---

[1] This interpretation was brought back to prominence and explored in detail by Leslie Ballentine; see L. E. Ballentine, *The Statistical Interpretation of Quantum Mechanics*, Rev. Mod. Phys. **42** No. 4, 358-381 (1970) and L. E. Ballentine, *Quantum Mechanics: A Modern Development* (World Scientific Publishing, Singapore, 1998).
[2] M. F. Pusey, J. Barrett, T. Rudolph, *On the reality of the quantum state*, Nat. Phys. **8**, 475–478 (2012).
[3] M. F. Pusey, J. Barrett, T. Rudolph, arXiv:1111.3328 [quant-ph], Nov 14, 2011.
[4] Valentini says: "… I think the word 'seismic' is likely to apply to this paper." E. Reich, *Quantum theorem shakes foundations*, Nature: News (Nov. 17, 2011).

The first thing to note, and it overarches all, is that, quantum mechanics (QM) *is a statistical theory*. The Born rule,[5] which is an integral part of standard QM, effectively states this. Indeed, one does not need a separate postulate stating that quantum mechanics is statistical in the primary meaning of the word, for one cannot do anything with one measurement in QM. One needs many measurements to generate a histogram of possibilities, i.e., a probability distribution. If formal QM had no handle connecting it to such observed distributions, being the statistical theory it is, it would evaporate, i.e., would not be a theory at all, as it would have no relation to reality. Furthermore, even when one is discussing only a few measurements such as in the GHZ theorem,[6] in the background are things of a statistical nature standing in the stead of the actual measurements---this is revealed by statements like "if we measured this we would get this or this." Hence, PBR cannot be not be talking about this primary meaning of statistical, but must be talking about a narrower one.

That said, there are five issues to discuss: **1)** Further expansion on the fact that the definition of statistical or "information carrying" that must be applied to bring PBR into play is too narrow (and not near exhaustive) to disprove any proper generic definition of "statistical interpretation" (and thus the ensemble interpretation). That is, even accepting the other assumptions of PBR, one is not forced to abandon theories that are in a proper sense statistical. **2)** Related to this, connecting PBRs' result to classical statistical ensembles (which has been done) can be a red herring, because there are ensembles used in classical statistical mechanics that do not meet the standards of the PBR definition so would therefore not be considered statistical! **3)** PBR needs to be considered in the light of the lessons learned from the analysis of Bell's theorem, as explained in various places.[7] **4)** QM does not have to be exact; that is, there could, for example, be other mechanisms in nature as yet not part of QM as we know it (such as the collapse theories of Pearle or Penrose).[8] Or, there could be a sub-quantum theory to which standard QM reduces in the limit, such as the equilibration (a kind of analogical thermalization) discussed by Valentini[9] in the de Broglie formulation of the de Broglie-Bohm interpretation. **5)** The statistical interpretation of measurement is not affected by the PBR theorem.

---

[5] This is the rule that $|\psi(x)|^2$ is the *probability* of finding the particle(s) between $x$ and $x+dx$.

[6] Or, in more complex systems, the Kochen-Specker theorem. The GHZ theorem is due to Daniel M. Greenberger, Michael A. Horne, and Anton Zeilinger.

[7] A. Rizzi, *The Science Before Science*, (SBS), (IAP Press, Baton Rouge, 2004). A. Rizzi, *The Meaning of Bell's Theorem*, arXiv:quant-ph/0310098v1. A. Rizzi, *Physics for Realists: Quantum Mechanics*, (IAP Press, Rochester, NY 2018).

[8] P. Pearle, *Collapse Miscellany* in *Quantum Theory: A Two-Time Success Story* edited by Struppa, Tollaksen, (Springer, NY; 2014) pg 131, Penrose, *The Road to Reality* (Alfred A. Knopf, NY, 2004).

[9] Anthony Valentini, *Inflationary Cosmology as a Probe of Primordial Quantum Mechanics*, arXiv:hep-th0805.0163v2, 1-44, Sep 9 2010. Valentini, *Cosmological Data Hint at a Level of Physics Underlying Quantum Mechanics*, Scientific American, Nov. 2013. A. Valentini, *Signal-locality, uncertainty, and the subquantum H-theorem. I*, Phys. Lett. A **156** No.1, 5-11 (1991). A. Valentini, Signal-*locality, uncertainty, and the subquantum H-theorem. II*, Phys. Lett. A **158**, 1-8 (1991). S. Colin and A. Valentini, *Mechanism for the Suppression of Quantum Noise at Large Scales on Expanding Space*, Phys. Rev. D **88**, 103515 (2013).

Before beginning an analysis of these issues, we summarize the PBR result[10] in the next section and, in the section after, we discuss, at a generic level, the primary model that we will use *both* to probe the nature of the requisite statistical analysis *and*, in particular, as an analogy to the statistics of quantum mechanics. In that section, we build both the model and the most general framework (and no more), for discussing QM. Then, we will go point by point through the above issues.

## Brief Summary of PBR Theorem

In this section, I will use simple language to reach the heart of the model and avoid possible confusion. My starting point, and thus my terminology, is different than that of PBR. This is to avoid the very confusions I am trying to clarify.

PBR begin by associating every quantum state with a physical state. The quantum state does not exhaust all possible knowledge of the physical state. A system prepared in a quantum state $|\psi\rangle$ is associated with a probability of producing a physical state $\lambda$, by a probability distribution, $\mu(\lambda)$. At the outset, it seems possible for the support of two non-orthogonal quantum states to have $\mu$'s that overlap, i.e. have common support.[11] We will discuss this in more detail later; for now, note that if they overlap the $|\psi\rangle$'s give less information than if they do not (something we also discuss later). They, then, show that, given the possibility of preparing two *physical* states without them physically influencing each other, such overlap contradicts the predictions of quantum mechanics. Now, we are ready to discuss the issues we raised above.

## QM and our Primary Analog, the Temperature of an Ideal Gas

The core of this paper requires deepening and clarifying our understanding of how we apply statistics to a physical system. In particular, we are trying to better understand how to interpret quantum mechanics when it labels a system by a quantum state (e.g., a ket such as $|\psi\rangle$). The quantum state is not to be identified with a single state of reality, such as the momentum and position of a particle, but it is, in this view, just what it appears to be. $|\psi\rangle$ labels an ensemble of similarly prepared systems.

We will start with things that we sense, as we finally always must. We take temperature. The temperature of this body being different than that one, is one way we can distinguish this one from that one. Of course, any such ordinary macroscopic body is composed of atomic parts. A simple case to analyze through looking at its parts would be an ideal gas in a fixed volume *V*. We know the temperature can distinguish one such box filled with an ideal gas from another. If we consider a somewhat thermally isolated box, we know the total energy (*E*) and the total number of particles (*N*) are fixed (micro-canonical ensemble). In appendix A, I show that the temperature is in one to one correspondence with the total energy and that it is approximately proportional to *T* for macroscopic *N*. (all other parameters remaining fixed). There is always such a correspondence even for small *N*; it is just that, in that case, it is not very useful for statistical analysis of the system in which temperature has meaning.[12] Still, this shows that *T* is characteristic of a system with fixed *E*.[13] This is not usually a direction we think; we usually think of *T* as representing the average energy of individual particle. Here we are trying to characterize the whole system. For this reason, though the derivation in the appendix might look at first glance to be totally standard, it is not because it approaches it from this different perspective, i.e. from using *T* to characterize the whole system as a substitute for *E*. This affects the way calculation is started, the direction it goes and how it is understood and thus makes it important to read and digest, as it is non-standard in this way.

---

[10] To understand the details of the technical argument, I recommend the PBR article itself (see footnote 2) along with the article on the web by Matt Leifer at: http://mattleifer.info/2011/11/20/can-the-quantum-state-be-interpreted-statistically.

[11] These concepts relevant to ontological models were introduced by N. Harrigan and R.W.Spekkens *Einstein, Incompleteness, and the Epistemic View of Quantum States*, Found. Phys. **40**, 125 (2010). They say "a hidden variable model is *ψ-ontic* if every complete physical state or *ontic state* in the theory is consistent with only one pure quantum state; we call it *ψ-epistemic* if there exist ontic states that are consistent with more than one pure quantum state." This language might be a chief driver in losing the larger understanding of statistical theory, for *ψ*-epistemic models do not constitute the sum total of all statistical models.

Now, there are two ways of thinking about an ideal gas coming to equilibrium.

In the *first*, we imagine that we put the *N* particles in the box in some random state except for the fact that we require that they have a *total* energy *E*. We also suppose that the gas molecules have some weak interaction between them so that the system can, over time, come to equilibrium.[14] That is, it has reached the point where it is most likely to be found in the energy binning state (see Appendix A) that has the most microstates.

The *second* way to look at equilibration is to imagine that we start by putting our box on a stove of temperature *T*. After the box has come to equilibrium with the heating element, we know the box is at the right temperature, and we take it off. However, we are not done. We only keep those boxes that have energy *E*. In this way, we have a preparation method that can produce an ensemble of boxes distinguished by their temperature *T* (and their energy *E*). Still, for a given temperature, any given box, nearly all the time, will be different than the next. This is the sense in which it is statistical; we *do not know its microstate just its macro-state* (which still includes many different microstates).

Note that in the language of statistical mechanics, for both cases, the ensembles are called micro-canonical. Now, we are ready to dive into our PBR discussion.

### What Types of Descriptions Fall under the Heading of Statistical?

Our <u>first two issues</u> are intimately connected. In their arXiv paper, PBR say: "*the statistical view of the quantum state is* that it merely encodes an experimenters information about the properties of a system. We will describe particular measurement and show that the quantum predictions for this measurement are incompatible with this view." (I added italics to point to the most important part of this statement for us at this moment.) They wisely excise these lines from the version published in *Nature*. In fact the word "statistical" is gone from the article all together. Indeed, B&R along with two other authors wrote a follow-on paper titled: *The quantum state can be interpreted*

---

[12] So, we might, for small *N*, choose to call it *β* as *T* connotes the statistical meaning that depends on large *N* and reflects the temperature of physical bodies. (see Appendix A for more discussion).

[13] The temperature characterizes the energy binning that is most likely for the system of fixed *E* and *N*. The concept of temperature viewed through the parts of a system, i.e. *T*, is a statistical one; it does not appear until one considers the most probable states.

[14] If we are willing to wait, we can just allow the finite mass of the box to allow the momenta (and thus the energy) to redistribute. We should also note that one must be careful to not to pick a special initial state that will not equilibrate. The need to do so is well known in such statistical analysis.

*statistically*.[15] Unlike how it may sound, it does not explain how the PBR theorem and statistical understanding can coexist. As we have already said, this must be the case. Now, in discussing the first two points we are trying to cut through the actual as well as possible confusion and finally get to what actually was proved by PBR.

### *"Minimal" Information: Classical Thermo and Quantum*

First, we explore what one means by "statistical" by discussing the first robust (and most common) example of statistical theory: thermodynamics. Generally, one resorts to statistical analysis when one is either lacking information about the system under study or when one wants to focus on one aspect not another, so that it is convenient to ignore part of the potentially available information. In thermodynamics, one typically just does not know what the state of individual molecules is; that is their momenta, angular momenta, positions etc. are not known. Even if we did know them, it is not what we are after. Our senses give us direct access, for example, to the temperature of a copper pipe. We are thus, not interested in the details of the inner structure of the pipe. We are interested in temperature, which is reflected in the general state of its parts. If I knew the detailed motion of the parts, it would be too much information for me to process and to relate back to the temperature. In fact, continuing to follow our example, it is known that the temperature relates to the statistical state of the atoms of the copper pipe. We only need and want that much information. It is not all of information; it is part of it. This is generally true of thermodynamics and the statistical mechanics that describes it.

Furthermore, to proceed, we have to leave even more of the particular reality out of account; we do this by making certain idealizations in order to make our analysis both conceptually analogically general and the analysis operationally simple at first.

Perhaps the simplest thermodynamic case is that of an ideal gas. Consider our ideal gas of temperature $T$ in a thermally isolated box of volume $V$ (which has pressure $P$), a micro-canonical case. Specifying this state does not specify the state of each particle of gas, and thus comes very far from specifying the complete physical state of the system. $T$, as discussed in the previous section, is a statistical variable in this sense; it labels the macroscopic system by specifying a certain *set* of allowed states (not a single allowed state), which we take to be equally likely to be populated. The treatment is clearly statistical. Indeed, the gas can be analyzed using statistical mechanics; assuming equilibrium is reached, one can derive, as done in the appendices, the probability of finding a particle with a given energy. Now, here is the interesting point, which is pivotal for the PBR definition of "mere" information.[16] The original PBR definition would say the treatment of the ideal gas is not statistical! The new PBR definition would say it is not "mere" information! Now, information means any knowledge; it could be complete or partial knowledge. It can never mean no knowledge. All $\psi$ does, *in any case*, is encode information about a physical system; there is no possibility that, for example, $|\psi\rangle$ is itself a physical thing. It is a symbol that refers to physical things. $\psi$ gives more or less information about a physical system. Information, informs you *of something* not nothing.

---

[15] Lewis, P. G., Jennings, D., Barrett, J. & Rudolph, T. *The quantum state can be interpreted statistically.* arXiv: 1201.6554v1 [quant-ph], Jan 31, 2012. Title changed when published in Phys. Rev: P. Lewis, D. Jennings, J. Barrett, T. Rudolph, *Distinct Quantum States Can Be Compatible with a Single State of Reality*, Phys. Rev. Lett. **109**, 150404 (2012).

[16] As well as the below quote in the text, *PBR* says, for example: "..then $|\psi\rangle$ can justifiably be regarded as 'mere' information."

And, a little information is not *mere* information, anymore than a little sugar is *mere* sugar, as it suggests when one has a lot of sugar that one has something besides sugar. Nonetheless, PBR says (in the Nature version of the article) "if the quantum state [$\psi$][17] merely represents information about the real physical state of a system, then experimental predictions are obtained that contradict those of quantum theory." This not a felicitous use of the word information; indeed, it is confusing. I propose one use the phrase "minimal information" (by this I only mean that it is minimal in the present context, i.e., relative to non-minimal.) What did PBR mean by "mere" information? And, why is the temperature of an ideal gas not mere information? Or, in my new language, why is it not "minimal" information and what, indeed, is minimal information?

Well, if one has two boxes with ideal gases in them, one at temperature $T_1$, the other at $T_2$, the states of the particles are distinct. That is, if we catalog the energy states of all the particles in the first box and then do the same for the second box, they will always be distinct, never the same.

One can see this by noting that the total energy of the *first* gas is (approximately[18]) $E_1 = NkT_1 = \sum_{i=1}^{N} \varepsilon_i$,[19] where $\varepsilon_i$ is the energy of the $i^{\text{th}}$ particle, similarly for the *second* gas. Thus, we have:

(1) $\quad\quad\quad\quad \varepsilon_1 + \varepsilon_2 + \varepsilon_3 + ... = E_1 \text{ and } \varepsilon_1 + \varepsilon_2 + \varepsilon_3 + ... = E_2$.

If graphed so that the $i^{\text{th}}$ axis of the *N*-orthogonal basis is the energy of the $i^{\text{th}}$ particle, these equation represent two parallel planes, so that none of their states overlap. This analysis emphasizes the point but it is not necessary, as it is already obvious that the energies in one system cannot all be the same as those in another, otherwise the states would add up to the same energy, belying the starting premise that systems are at different temperatures. Thus, the temperature distinguishes the state of reality of one box from another (but it is still statistical, as the details of the particles are not known). A different method of distinguishing one system from another[20] might not do this, so it would give less information. Thus our temperature method does not give "minimal" information.

This is an analog of what PBR says for a quantum state. They say that *a quantum state gives mere information ((relatively)minimal information) if one cannot in principle deduce from the physical state (all the known and unknown properties) which quantum state the system is in*.[21] So, in a "non-minimal" case, the quantum state does not tell you what the entire physical state is, but it tells you which ones it isn't.

---

[17] Another way to read "quantum state" is as: "the physical states prepared according to $\psi$"; this gives the same problem but has the additional issue of referring to physical reality as "representing" something.

[18] Again, the important general point that arises out of the analysis in the appendix is that, though this is an approximate result: for a given temperature there is always only a single physical value for the total energy establishing the correspondence between temperature and total energy.

[19] This is the classical continuum limit result calculated in the appendix. Using quantum mechanical energy bins, in one dimension one gets: $E=kT$; in 3D, one gets: $E=(3/2)kT$.

[20] As an example where the distributions overlap, consider the case of two isolated ideal gases each at the same temperature and pressure, but in boxes of different volumes $V_1$ and $V_2$. The ensembles labeled by $V_1$ and $V_2$ would overlap. Yet, one clearly has some information (but not all the information, and it is clearly a statistical theory).

[21] Said another way, PBR asks whether each distinct $\psi$ refers to a group of distinct physical states or not.

To start their analysis, they take each quantum state to correspond to a set of physical states; initially no assumption is made about whether there is overlap between the physical states so associated. Consider say $|\psi_1\rangle$ and $|\psi_2\rangle$, PBR takes the first and second to be associated, respectively, with the distributions $\mu_1(\lambda)$ and $\mu_2(\lambda)$, where $\lambda$ represents a group of physical states to which one does not have direct access (analogous to the momentum of the particles in the ideal gas case). [22] Indeed, we will see that there are two ensembles in play, a larger one from which one picks a member $\lambda$ (and for which $\mu$ gives the probability) which itself labels a second sub-ensemble of physical states from which one will pick when carrying out a measurement. This will become clear in this subsection.

To get an example a situation where $\lambda$ represents such a sub-ensemble of a larger ensemble, return to our ideal gas of total energy $E$. Here, the particles can be in any state in which the total energy adds to $E_1$. And, I do mean *any* such state, even after equilibrium is reached. Even after equilibrium is reached, all the microstates of the system are accessible. To digest and deeply penetrate the meaning of this point and to approach a statistical analysis of physical systems in general from our unique perspective, we analyze our $N$ particle ideal gas in detail in appendix A.

Notice, from the derivation in the appendix, that equilibrium (i.e., the condition in which the gas has already "explored" the possible microstates enough to have found its way into the energy-binning state with the most microstates) only implies that it is highly likely that the individual system particles are in the individual energy states with the most available microstates (this is an energy bin-defined state in which, for our case, the number of particles with energy $\epsilon$ is proportional to $e^{-\epsilon/kT}$). It does not mean that any given member of the ensemble is in that configuration. After preparing our ideal gas at temperature $T_1$, it will be in one of many allowable states, which we schematically illustrate below on the left in FIG. 1. The same will be true for the preparation of our second box at temperature $T_2$; this is illustrated on the right. The shape of the distribution is only meant to be suggestive that there are some states that are much more likely to be picked out of the ensemble of possible states (the actual distributions have sharp delta-like peaks). Note how the states in each ensemble are non-overlapping[23]; that is no member of one ensemble is also a member of the other.

---

[22] If the μ's have overlapping support, the situation is said to give "minimal" information; otherwise we say it gives non-minimal information, i.e., more information. In PBR's language, one says: it is not "mere" information, not epistemic.

[23] This would also be true for two boxes of ideal gas in thermal contact with finite sized heat reservoirs of different temperatures. While the ideal gases would no longer confined to be a certain energy and thus the argument for distinct physical states given for the isolated box would fail, it would not fail for each box+heat-reservoir system as each of these systems is isolated. So including the full reality gets one back to non-overlapping ensembles.

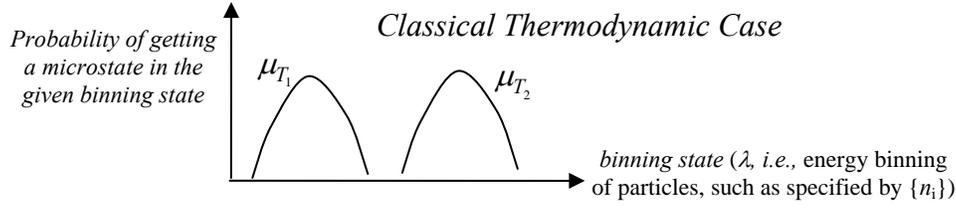

FIG. 1: A graph of the binning state, $\lambda \equiv \{n_i\}$ or ($\rho_E$ in continuous case), called an intermediate state in the appendix, *versus* the number of microstates available for the given binning prescription $\{n_i\}$. The number of microstates is proportional to the probability. Note there is a binning prescription that has the greatest number of microstates. This multiplicity arises from the ability to switch particles without changing the overall number of particles in each bin. Switching particles does not change the binning state, but does change the physical state of the system, because different particles are in different places. *The binning state is an ensemble of individual particles states (microstates); that is, each member of a binning state is a microstate. The macro-state, specified by the temperature, is an ensemble of binning states.* We label each binning state by $\lambda$ to make an analogy to PBR's $\lambda$ used to analyze QM.

In the case of a quantum mechanical system in the state $|\psi_1\rangle$, the probability of finding the system in state $\lambda$ is $\mu_1(\lambda)$, so we get the graphs shown here.

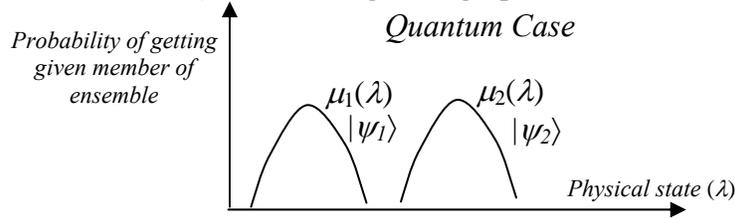

FIG. 2: Probability distributions for two quantum states.

We introduce a second function, $\xi_k(\lambda)$. It is a function of each possible ensemble (labeled by $\lambda$); it is universal in that it does not matter how the ensemble from which it was sampled was prepared. That is, stating it more formally, it does not matter which $\mu_i$ that $\lambda$ is in the support of; $\xi_k(\lambda)$ is the same. $\xi_k(\lambda)$ gives the probability that the physical state specified by $\lambda$ (which can represent an infinite number of parameters) will, when subjected to a measurement of some observable associated with the eigenstates $|k\rangle$, yield an outcome $k$. In this way, since the probability of getting the given measurement is independent of the chances of getting $\lambda$ from a system prepared in a given state, say $|\psi_1\rangle$, we get:

(2) $\quad |\langle k|\psi_1\rangle|^2 = \int_{\Lambda_1} \xi_k(\lambda)\mu_1(\lambda)d\lambda$, where $\Lambda_1$ is the support of $\mu_1$

To better understand the elements of this statement; notice the following, for a given $\lambda$, the outcome must be one of the values of $k$. This means we must normalize the probability by requiring: $\sum_{k=1}^{\infty} \xi_k(\lambda)=1$, or in the continuum case: $\int \xi_k(\lambda)dk=1$. Similarly, a given $\mu$ gives the probability of getting a member $\lambda$ of the ensemble described by $|\psi_1\rangle$, so it is normalized separately as: $\int_{\Lambda_1} \mu_1(\lambda)d\lambda=1$.

With these details in place, we give the requisite definition more completely: quantum mechanics contains minimal ("mere") information if some $\mu_i$ have common support of non-zero measure. Or in said in the negative, quantum mechanics is *not* conveying minimal information *if any given nonzero measure of states of λ, call it* $\Lambda$ *, only appears associated with one quantum state*. We write this as:

(3) $\qquad$ QM is not minimal information if $\Lambda \to |\psi\rangle$

For concreteness, suppose we are measuring the *energy* of a *single-particle state* $|\psi\rangle$--- so that $|k\rangle$ is an energy eigenstate, where $k$ is associated with an energy (eigenvalue) of the system which we will now refer to as $E_k$, so that $k$ can count the energy state under consideration. We can take the eigenvalue to give the average energy of the particle, where the average is taken over the ensemble of possible states as shown in equation (2).

Now how do we compare this with our ideal gas? In particular, what measurement shall we take for our ideal gas to be analogous to this and, in particular, to enable us to write a $\xi$? We measure the energy of one the particles. Say, in every microstate, we have painted one of the particles red.[24] This allows us to ask what its energy is. The answer will depend on which microstate. Even giving the binning state (intermediate state), i.e. $\lambda$, does not specify this energy. We use $\xi_\epsilon(\lambda)$ to give the probability of measuring the particle to have energy $\epsilon$. This is our analog.

Remember $\lambda$ is *not* a complete specification of our system; indeed, we need further freedom in order to have a $\xi$ that is a nontrivial probability distribution.[25] As mentioned earlier, $\lambda$ labels which sub-ensemble (which binning state) out of the larger ensemble of binning states that each yield a total energy $E$.

For a given state $\lambda$ in a box prepared at $T$, there is a certain probability, $\xi_\epsilon(\lambda)$, of getting energy $\epsilon$. So the total probability, which is obtained by running the measurement over a large sample of boxes prepared at $T$ is given by equation:

(4) $\qquad P_T(\epsilon) = \int_{\Lambda_1} \xi_\epsilon(\lambda) \mu_T(\lambda) d\lambda$

In fact, because there is a sharp peak (again, not indicated in the Figure 1) in $\mu_T(\lambda)$ at $\lambda = \lambda_0$ (which corresponds to the binning the particles according to the exponential in energy), we have: $P_T(\epsilon) \approx \xi_\epsilon(\lambda_0)$. By repeating the experiment many times, we can determine the probability of getting energy $\epsilon$.

Clearly, unlike what might be gleaned from some of the statements about minimal information states, such systems yield a viable statistical setup. Let me emphasize, this is so despite the fact that the $\mu$'s for our two different temperature boxes do not overlap. Giving the temperatures doesn't exhaust what is knowable about the system, even in the abstracted form that we have reduced it to (i.e., an ideal gas). However, it gives more information than if the method of labeling the system were "minimal."

---

[24] "Painting one particle red" is just a way of indicating that we can distinguish it from the rest and then measure its energy state. Whatever method of distinguishing our chosen particle that we choose, it should not affect the dynamics of the particle at our level of analysis. For example, considering one of the particles to be in an excited states, under certain conditions, would enable us to distinguish it without changing the nature of the system (at our level of analysis).

[25] Remember because probability is a result of ignorance (i.e., it results from ignoring, leaving out, part of a system), we cannot completely specify all elements without ceasing to have probability distributions. PBR say, in the first arXiv version of their paper (see footnote 3): "If the quantum state is statistical in nature…, then a full specification of $\lambda$ need not determine the quantum state uniquely." In fact, if it is statistical in nature, it cannot be uniquely determined.

### *On the levels of Information*

Before continuing, note that even if the domain of support of each prepared ensemble is identical, that is complete overlap of the supports, we still have a statistical theory. Indeed, perhaps this is the absolute minimal information type of theory?

As a concrete example of *complete overlap*, take the case of two differently prepared ensembles labeled $A$ and $B$. Suppose we are considering a system composed of 100 marbles; there are green, red, blue and white marbles. Ensemble $A$ is defined by always putting 25 of each color, while $B$ has 50 green, 15 red, 15 blue and 20 white. By sampling the ensembles, I can determine something about the system, but not, of course, what the color of the one I pick will be; that is only given probabilistically, which I can discover by experiment.

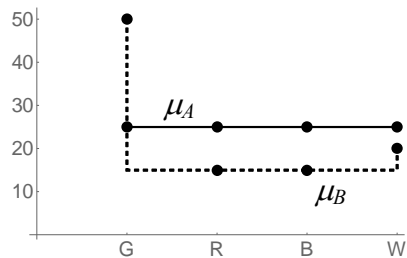

FIG. 3: Example of point probability functions for ensembles of marbles $A$ and $B$ that completely overlap. The solid line indicates the large dots that belong to $A$ and the dotted line indicates those for $B$. There are four different colors of marbles: green (G), red (R), blue (B) and white (W).

Next, consider the *non-overlap case*. Suppose I sample, (not going through all the marbles just some) a group of bins labeled $C$, and I find that 25% of the time I get green, 75% red. And, I find for another group labeled $D$ that 50% of the time I get blue and 50% white. Further, suppose I know, somehow say from a PBR-like proof, that the domain of supports do not overlap. I have more knowledge than in the previous case; I know that the particles in one bin will not contain certain colors. For example, getting a green in one bin will tell me that I will not find green in the other bin.

I can also have *partial overlap*. For example, suppose I'm told that group $E$ has 50 green, 25 red and 25 blue, while group $F$ has 25 red and 75 white. Supposing I was told the distribution for the other two cases as well, this gives me an amount of information between that of the other two. If I get a green marble, I know it must be from group $E$, unlike the total overlap case. If I get a red marble I don't know whether it came from $E$ or $F$, unlike the non-overlap case. This *partial overlap* case[26] is what one gets if one has a two ideal gases, each at the same total energy, $E_1$, but differentiated as follows.

---

[26] The following example is a special case of partial overlap in which one is a subset of the other in which not all of its boundary is inside of the other. One could imagine a case where one is completely surrounded by the other, so it is kind of an island (the smaller set) inside of an ocean (the larger set). Even this would still be considered partial overlap as picking a point in the "ocean" will distinguish the larger set.

One has zero as its lowest possible energy, while the other has $\varepsilon_0$ as its lowest energy.[27] This case of non-zero ground state is discussed in Appendix A2.

Notice that each of these categories of statistical information gives information, but the most information is given by the non-overlap category.

### *Conclusion of Statistics discussion*

We have shown the range of things that are properly statistical and the definition of "mere" (minimal) information invoked by PBR does not reach the ensemble interpretation. In particular, we showed that the ensemble view used in an ideal gas would not fall under PBR's definition! Having laid out the premises of the theorem, i.e. the assumptions inherent in the definitions,[28] we are ready to address the first assumption that is called out by PBR as such.

## **Some Lessons of Bell's Theorem** (*3rd issue*)

As mentioned earlier, to accomplish their proof, PBR requires that the two distributions, $\mu_1$ and $\mu_2$, associated with two orthogonal states (say $|\psi_1\rangle$ and $|\psi_2\rangle$), must be prepared independently of one another. This assumption is intimately related to Bell's theorem, in so far as that theorem, modulo some other assumptions,[29] indicates that there are super-luminal interactions implied by QM.

Bell's Theorem if analyzed deeply points to important general distinctions that need to be made in discussing quantum mechanics at its core.[7] These, in turn, are generally in the statistical realm and can be missed if one is not attentive to the unique nature of statistical analysis. For example, as explained in the references mentioned in the introduction,[7] to do the proof, Bell has to consider two non-compatible measurements at the same time; because of this, the possibility opens up[30] that the very nature of QM is that that the particle interactions with a detector sometimes result in no registered signal.[31,32] A relatively recent experiment[33] (that uses a radically new approach called

---

[27] They will have different "temperatures," but here we distinguish the macro-states by total energy.

[28] I say "premises" because they can also be considered as assumptions of the theorem though they are implicit, appearing in the form of definitions. They come before the actual stated assumptions.

[29] Including ones not often understood clearly.

[30] Another such example occurs in the GHZ proof. If one recognizes the statistical nature of QM, then one quickly sees that one has to consider several alternative potential measured values simultaneously that are not, in fact, simultaneously measurable. This leads, for example, to the widely recognized need for QM contextuality. Moving to a general point that we have implicitly alluded to before, it is not that "no-go" proofs are not important, but that they often reveal more than their simple conclusions; in particular, their explicit and hidden assumptions reveal a lot. Bell famously said, I think in trying to bring out this point: "…long may Louis de Broglie continue to inspire those who suspect that what is proved by impossibility proofs is lack of imagination" J.S. Bell, *On the Impossible Pilot Wave,* Found. Phys. **12** No. 10, 989 (1982).

[31] Similar caveats, at some level, probably also apply to PBR reasoning.

[32] This caveat can be lumped under the title, "detection loophole," but notice the title is not appropriate in so far as something that is in the nature of quantum mechanics as this mechanism would be is not a "loophole." Loophole tends to imply one just has not yet found a good enough (efficient enough) detector and if one did then we would see that quantum mechanics is correct in its predictions. In the hypothetical (that seems not to be the case), quantum mechanics would *continue* to be verified with better and better detectors, because such experiments would keep giving the same results for the detections observed.

[33] B. Hensen et al, *Experimental loophole-free violation of a Bell inequality using entangled electron spins separated by 1.3 km,* Nature, **526**: 682-686 (2015)

heralding) seems to have ruled this out, making superluminal[34] travel a necessity.[35] With super-luminal interactions considered as integrated into the core of quantum mechanics, there is no reason to suppose that one can independently prepare the systems without a lot of care. Indeed, if the action is non-local in the sense of instantaneous, it would be natural to expect that no amount of care could accomplish an independent preparation.[36] The latter would pose a principled block to PBR's conclusions, the former would put an effective block.

## Is Quantum Mechanics the Final Word on Reality?

*The fourth issue*, which affects how we situate the PBR theorem in our thinking, is the theorem's relation to sub-quantum theory, or generally the reality which quantum mechanics describes. Even if one knew that quantum mechanics represents the absolutely best *possible* formal theory of the generic nature of physical reality, not just the best we have right now, then one would still want to know what it described. PBR are in fact doing just that, i.e. trying to delineate exactly what QM means.[37] In fact, it would not be prudent to think we are done, holding that QM is the last word. Many physicists have done productive work in this area of the foundations of quantum mechanics, with their output affecting others areas of physics, including influencing areas one might not expect to be influenced.[38]

Bohmian mechanics, Nelson's stochastic mechanics and the walking droplet analogy are some of the most important direct fruits of this effort. Bohmian mechanics, for example, looked at from the de Broglie viewpoint introduces the possibility of a non-equilibrium state where the Born rule does not apply. In that regime, results not predicted by QM would be measured. Valentini has analyzed this possibility in some detail, discussing the possibility of an early age of the universe before the equilibrium was established in which standard QM would not apply.[9] These theories not only remind us of the already obvious fact that the physical world QM describes is there, but make clear the possibility of successfully exploring the depth that supports the evident beauty of quantum mechanical theory (so called sub-quantum theory). They help manifest the reality of the possibility of obtaining more information[39] about the quantum mechanical aspects of the world. Moreover, they give real, though partial and sometimes confused, insights into the nature of that world.[40]

---

[34] Of course, such superluminal travel would have to be unmeasureable to avoid violating special relativity; said another way, to avoid violating special relativity, one must, in principle, not be able send information with the superluminal interaction.

[35] Assuming that one keeps the real world, upon which, it should be obvious, the whole of physics depends.

[36] The preparation independence assumption is discussed in the following: M. Schlosshauer, A. Fine, *No-Go Theorem for the Composition of Quantum Systems*, Phys. Rev. Lett. **112**, 070407 (2014), S. Mansfield, *Reality of the quantum state: Towards a stronger ψ-ontology theorem*, Phys. Rev. A **94**, 042124 (2016).

[37] "But our present QM formalism is not purely epistemological; it is a peculiar mixture describing in part realities of Nature, in part incomplete human information about Nature-- all scrambled up by Heisenberg and Bohr into an omelette that nobody has seen how to unscramble. Yet we think that the unscrambling is a prerequisite for any further advance in basic physical theory. For, if we cannot separate the subjective and objective aspects of the formalism, we cannot know what we are talking about; it is just that simple." E. T. Jaynes, *Complexity, Entropy, and the Physics of Information*, (ed. by W. H. Zurek) 381 (Addison-Wesley, 1990). This is quoted in reference 2.

[38] For example, the blossoming field of quantum computing is intimately connected with thought about the nature of entanglement (to which Bell is a primary contributor).

[39] Indeed, we would want to find a formal theory to describe the sub-quantum world.

In addition to being interesting in themselves, the proposed collapse mechanisms mentioned in the introduction point to how pregnant quantum mechanics is to lead us to unknown territory. In 1980, Hawking declared physics could be over soon,[41] but he later regretted and reversed the statement. Since the late 1800's, many have thought physics *was* already effectively over.[42] We don't want to keep making this mistake. We must have strong reasoning and evidence that we have the *most comprehensive, complete theory possible* before we say we have such a "final" theory; indeed, at this moment in history, very few would say we already have a final "theory of everything."

If we explore the meaning of QM in an open way, we can see points like that of S.M. Halataei's page-long comment[43] to PBR's article in which he says: "The conclusion would be accurate if there was no minimum for $\varepsilon$. However, if, due to practical limitations, there is a minimum for $\varepsilon$ ($\varepsilon_{min}$) that cannot be passed for any $\theta$, then the test would not be very informative." In other words, even if one accepts the PBR assumptions, it could be that, as a practical matter, one cannot ever be forced to say that there is no overlap (i.e. the information is minimal or "mere"). More importantly, one can take this another step and say it could be in the nature of physical reality to work in such a way that such measurements are not possible. This again, would indicate the presence of something beyond quantum mechanics, because, as far as we know, QM has no such block.

Indeed, PBR respond by defending QM as the last word: "There is — according to standard quantum theory — no fundamental limit on how small $\varepsilon$ can be in a careful enough experiment. Hence, any model with overlapping probability distributions for a distinct pair of quantum states makes different predictions from standard quantum theory, and such a model could be ruled out in a suitably sensitive experiment"[44]

## Measurement

The *last of our five issues* addresses PBR's mention of wavefunction collapse toward the end of their paper. Their statement is easy to interpret as implying that the ensemble interpretation is put in jeopardy by their conclusions.[45] It is not clear what argument would be made in this regard, especially given the fact that dBB stands as a counter example. It is a statistical interpretation.

---

[40] See footnote 7 for reference to *Physics for Realists: QM* in which those insights are joined with others and brought into a coherent whole pregnant with the possibility of more investigation to further deepen our understanding.

[41] Talk titled: *Is the end in sight for theoretical physics?*, given at his inauguration to the Lucasian Professor of Mathematics at University of Cambridge. Former holders include Newton and Dirac. He said: "I want to discuss the possibility that the goal of theoretical physics might be achieved in the not too distant future, say, by the end of the century. By this I mean that we might have a complete, consistent and unified theory of the physical interactions which would describe all possible observations."

[42] See "The Last Word in Physics" from S. Jaki, *Patterns or Principles and Other Essays* (Intercollegiate Studies Institute, Delaware, 1995) for examples (involving top flight scientists) of this tendency in physics history for scientists to think physics is finished.

[43] S. M. H. Halataei, *Testing the Reality of the Quantum State*, Nat. Phys. **10**, 174 (2014).

[44] PBR also say in their response: "Difficulties with the finite precision of experiments are unavoidable in physics. There will always be an infinite number of theories that predict the same results to within current experimental error. There is always the hope that future experiments will be able to distinguish them, and in the meantime we can use other criteria, such as simplicity, to select a preferred theory. We join with most physicists in preferring quantum theory at present."

To proceed, we note that there is a "layman's" description of PBR[46] on the web that may give insight into what is being thought here. It is written, incidentally, by a philosopher defender of the so-called transactional interpretation of quantum mechanics. As currently there is no clear statement as to how a statistical understanding of collapse, which is not itself a change in the wavefunction,[47] is put in jeopardy if one accepts the PBR conclusion, we will use the understanding presented in this article to create an argument that someone might make against the ensemble interpretation.

In the article, collapse is likened to trying to find a piece of hardware, say an odd-sized bolt that you've heard is available only in one store in the city. You collapse the wavefunction by calling a nearby "home improvement" store and finding that they have it. This is likened to a measurement that gives you information. Before the bolt might have been in any one of many stores. Now, you know that it is in this much smaller region occupied by the store. You further collapse the wavefunction by going into the store and seeing the hardware section. Finally, you ask for help and find that it is in aisle #16. Lastly, you do one final measurement and find it half way down on the aisle near the floor on the right. Each one of the descriptions (wavefunctions) included the other. There was overlap in the support of the probability distributions. But, PBR, if one accepts their premises, has shown that such probability distributions would lead to predictions that disagree with those of quantum mechanics, therefore the statistical view of collapse is invalid.

What is wrong with this? There is an artificial restriction introduced in the above argument. The restriction is implicit, but it is there. The argument assumes that the measurement does not affect the system in any way. By contrast, in quantum mechanics when a measurement is done, the object that we want to measure interacts with a device to measure it (leave aside the environment for simplicity). In the end, the two are, generally, entangled after the measurement. This means a new quantum state exists that did not exist before. And, by the PBR theorem itself, because the system is in a different quantum state, it will be physically different than the state of system (object +the measurement device) before the measurement.[48] Here again the word statistical is used in a funny way, outside of its generic meaning. The argument assumed that a statistical analysis must assume the measurement does not affect the system. If we let the system change,[49] then the attack on the ensemble interpretation of the above argument evaporates. That is, one gets along fine without requiring that the probability distributions overlap.

---

[45] They say: "If there is no collapse, on the other hand, then after a measurement takes place the joint quantum state of the system and measuring apparatus is entangled and contains a component corresponding to each possible macroscopic measurement outcome. This would be unproblematic if the quantum state merely reflected a lack of information about which outcome occurred. However, if the quantum state is a physical property of the system and apparatus, it is hard to avoid the conclusion that each macroscopically different component has a direct counterpart in reality."

[46] R. E. Kastner "Why quantum theory isn't a shell game (PBR Theorem for the layperson)" https://transactionalinterpretation.org/2014/06/14/why-quantum-theory-isnt-a-shell-game-pbr-theorem-for-the-layperson/

[47] By "collapse," I mean the general understanding of measurement of a given interpretation. In the ensemble interpretation, there is no collapse in the sense of requiring the reduction of the wavefunction to an eigenstate of the measured variable. In the context of the ensemble interpretation, the word collapse just refers to the process of discovering what value the variable has for a given measurement.

[48] This is not at all like the hardware store analogy; in that analogy the bolt would be moved somewhere else every time you probed to get information about its location-- clearly, this is not a good analogy for QM. A key reason is that the analogy treats the reality of the state as if it were one simple "there or not"

A very direct way to see that the ensemble interpretation is not damaged by PBR is, again, to note the possibility of the de Broglie/Bohm interpretation, which is equivalent to QM in its predictions and has no special collapse mechanism. It proposes that one member of an ensemble of physically distinct possible systems (different by their initial positions in configuration space) is created when we prepare a state. This is clearly statistical, clearly an ensemble interpretation.

## Conclusion

The confusion in the terminology and probably even in the understanding of the breadth of ensemble interpretation itself can lead one to think that PBR conclusions threatens or even cuts off the ability to hold that interpretation. We have shown that this is not the case, as the use of statistical ensembles already in classical thermodynamics shows the irrelevance of the overlapping probability to the ability to create such ensembles. In short, there are *implicit* false assumptions in the definitions that one can easily make if one is not careful with the meaning of the words one uses in this context. We have also shown that the explicit assumption necessary to get the PBR conclusion are called into question by lessons learned through Bell's theorem.

It is also clear that PBR's conclusions are contingent on quantum mechanics predicting exactly what nature does in all domains. We have no guarantee that this is the case; indeed, something has to give if the quantum theory of gravity arises. It seems likely it's *both* general relativity and quantum mechanics. More to the point, we should not assume physics is done, as that is the road to stagnation. Also, history shows a string of such false predictions.

Lastly, we saw how the statistical treatment of "collapse" survives even if one accepts PBR's conclusions. To bring this point home, we end by showing this for the paradigmatic case of Schrödinger's cat. In particular, we show that non-overlapping states do not imply the existence of two Schrödinger cats: one dead one alive. Sometimes one gets:

1) a cat in a changed new physical state but yet still alive along with an un-decayed excited particle also physically (ontologically) changed by the interaction in some way but not decayed.

2) and sometimes you get a dead cat (thus completely changed) physical state along with the particle that has decayed.

That is, consider the following very simplified cat experiment:

Assume I initially have a cat described by the quantum state |cat+> and an excited atom, |atom-e>, in the cat's brain. This atom can interact with the brain and decay (to a quantum state described by |atom-d>), to emit an X-ray that can kill the cat, |cat->

---

there as opposed to many entities that can be in many different states some of which stay the same as the states change. In a complex system, like our ideal gas, changing one element, would force us to say the system had a different $\lambda$.

[49] As we expect it would already, before detailed analysis, from our intuition about quantum mechanics.

However, it can also interact in such a way that the atom continues on its way out of the cat. Before the measurement (i.e., before atom-cat Hamiltonian evolves the state), I take the quantum state to be the (non-entangled) direct product:

(5)                              |atom-e>|cat+>

After the measurement (after interaction with cat's brain), we have:

(6)                              |cat+>|atom-e>+|cat->|atom-d>

If one accepts PBR's conclusions, then quantum state (5) and (6) also have no physical common support.[50] This simply means, because they are different quantum states, that the atom/cat system cannot have all of its physical properties the same before and after the interaction; this is fundamentally the scenario laid out above. In short, the physical states are changed by interaction; one cannot leave out what the effect of the measurement interaction.

We end by emphasizing the crucial point relative to this cat experiment, one that is easily missed. One might take the quantum state |cat-alive> to reference all the possible physical (ontic) microstates of the cat being alive. In such a case, the physical states associated with |cat-alive> must overlap with those associated with the quantum superposition state (e.g., ( a |cat-alive> + b | cat-dead>) ). Why? Otherwise, there would not be member of the ensemble for which the cat is alive when the system is in the quantum superposition state. And, this obviously contradicts quantum predictions. And, the overlap contradicts PBR.

However, it should be clear that it is not necessary to take |cat-alive> to exhaust all possible physical microstates of the cat being alive. Indeed, one cannot take the domain of ontic (physical) microstates states of associated with the superposition of two quantum states[51] to be simply the union of the physical microstates available to each of the two states of the sum. The superposition contains different physical microstates states than the union of the two separate states. The "+" in the superposition statement implies the introduction of new physical states. This is clear in the dBB interpretation where the superposition yields a pilot wave that is different than that of either quantum state (taken individually) that makes up the superposition. The figures below help relay this point. The first figure shows the ontological (physical) microstates for the quantum state |cat alive> and next to it, for the quantum state |cat dead>. Notice that these states do not overlap.

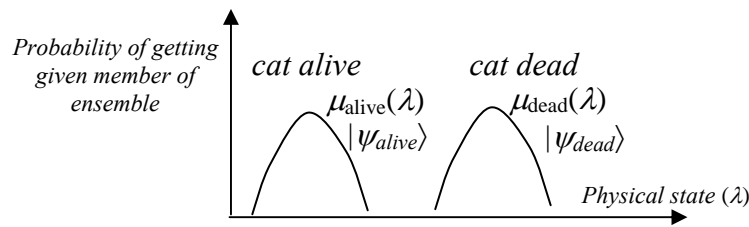

---

[50] Of course, |cat+>|atom-e>+|cat->|atom-d> is a different quantum state from both |cat+>|atom-e> and |cat->|atom-d> and so none of these have common physical support, i.e. none of the systems selected from any two of these ensembles will be the same.

[51] Said another way, the physical microstates that correspond to the quantum state (the "macro-state").

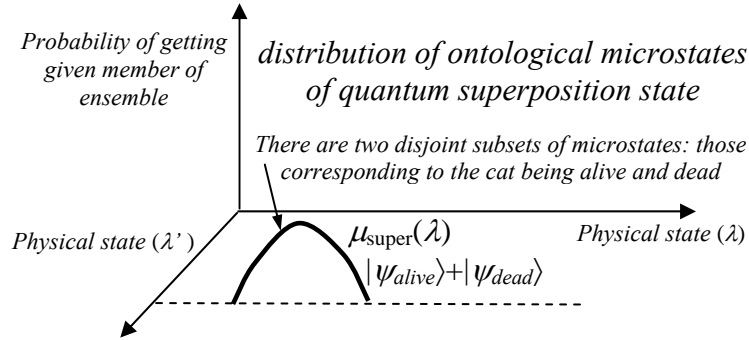

The last figure (immediately above) shows the ontological microstates for the quantum superposition state (|cat alive>+|cat dead>). Notice that the superposition has generated a whole new dimension of ontological states, which I label by $\lambda'$. I take the $\lambda'$ axis to be orthogonal to make clear the radical distinction that superposing states introduces. Of course, the physical states which correspond to the cat being alive and the cat being dead cannot overlap within this distribution; note, no attempt has been made to show how those states are distributed within the support of $\mu_{super}$. Like the micro-canonical ensembles of statistical mechanics, this superposition state has no overlap with the other states.

## Appendix A: Derivation of Statistics of Ideal Gas
($N$ particles, total energy=$E_1$ at temperature $T$)

(Note, on a glance, this title and the appendix itself might look like simply a standard derivation. However, it is not. It is written to address the issues raised by the PBR theorem about the nature of the statistical analysis of a physical system. Parts of the calculation are standard; however, it is approached and analyzed from a wholly unique perspective that, as mentioned in the text, affects the understanding, direction and final output of the calculation. Even the standard parts, in addition to being integral to the presentation, are needed to enable reflection on (and reference to) the key aspects of a statistical description of a physical system.)

As is evident from equation (1), there are many ways to assign the energy of each particle so as to get the total energy $E_1$, indeed infinitely many ways. To avoid the infinities, start with energy bins of finite size; later, we can take the continuum limit. Divide the energy axis from zero to $E_1$ into $M$ bins, and label the energy of a bin by the energy of the middle of the bin. Thus, the $i^{th}$ bin has energy $\epsilon_i$.[52] We take $M$ to be very large and we take $N \gg M$. Now, taking $n_i$ to be the number of particles in the $i^{th}$ bin, the energy and particle number constraint are:

(7) $$\sum_{i=1}^{M} n_i \epsilon_i = E, \qquad \sum_{i=1}^{M} n_i = N$$

Any possible way of arranging the particles that satisfies these constraints is a possible physical state for the gas of temperature $T$. We call each possible state a *microstate*. However, we divide these states up into classes of intermediate states. These intermediate states are simply defined by how we bin the particles. We specify the binning by listing the number of particles in each energy state. Thus, each intermediate state is defined by a list that looks like: $\{n_1, n_2, n_3 ...\} \equiv \{n_i\}$. Clearly, there are many different arrangements of particles that can satisfy any given binning assignment. Indeed, combinatorics shows that the number of microstates, $\Omega$, in a given intermediate state, $\lambda \equiv \{n_i\}$ is given by $\Omega(\{n_i\}) \equiv \dfrac{N!}{n_1! n_2! n_3! ... n_M!}$. The differences between the various microstates that make up an intermediate state are ignored. After all, the difference between microstates within one intermediate state just arises by switching the particles with each other. One can change which particles are where without changing the number in each bin. For example, we could switch particle 2400 in $n_1$ with particle 22 which is currently in $n_{49}$ without affecting the *number* of particles in each bin. This makes no difference, because, at our level of analysis, a particle with a given energy acts the same as another particle with the same energy. Thus, an ensemble of microstates makes up an intermediate state. We will not attempt to completely specify this ensemble, but we will need the number of states in it, i.e. $\Omega(\{n_i\})$.

---

[52] Note that our use of $\epsilon$, not $\varepsilon$, here; in the main text of the article, the $\varepsilon$ was the energy of a given particle in the gas; $\epsilon$ represents the energy of a given bin when a single particle is in it.

Now, many intermediate states can be assigned to the same macro-state, i.e. something evident on ordinary human scales. That is, the differences between the various intermediate states are not evident on that ordinary scale. Hence, we can consider an ensemble of intermediate states to represent a given macro-state, such as our gas with energy *E*. When we look in further detail at a given macro-state, we can find *any* member of this ensemble. This is what triggered our trek to understand the statistics of an ideal gas. But, we are not yet done.

Which member of the ensemble is most likely to be picked. Answer: the one with the most number of microstates associated with it. Thus, we need to maximize $\Omega(\{n_i\})$ with respect to $\{n_i\}$. Because an additive quantity is easier to work with than a muliplicative one, we define the entropy, *S* as:

(8) $$S \equiv k \ln \Omega(\{n_i\})$$

Now, using Lagrange multipliers to incorporate the constraints from equations (7), our maximization condition ($\delta S = \sum_{i=1}^{M} \frac{\delta S}{\delta n_i} \delta n_i$) becomes:

(9) $$\delta S = k \delta \left( \ln N! - \ln n_1! - \ln n_2! - \ldots - \ln n_M! - \alpha \sum_{i=1}^{M} n_i - \beta \sum_{i=1}^{M} n_i \epsilon_i \right) = 0.$$

Using Sterling's approximation ($\ln m! \approx m \ln m$, for large $m$)[53] by taking $n_i$ large and

$$\delta N = \delta \sum_{i=1}^{M} n_i = \sum_{i=1}^{M} \delta n_i = 0, \qquad \delta E = \delta \sum_{i=1}^{M} n_i \epsilon_i = \sum_{i=1}^{M} \epsilon_i \delta n_i = 0 \text{ gives:}$$

(10) $$\delta S = k \delta \left( N \ln N - \sum_{i=1}^{M} n_i \ln n_i - \alpha \sum_{i=1}^{M} n_i - \beta \sum_{i=1}^{M} n_i \epsilon_i \right) = 0$$

so: $\left( -\sum_{i=1}^{M} \delta n_i \ln n_i - \sum_{i=1}^{M} \delta n_i - \alpha \sum_{i=1}^{M} \delta n_i - \beta \sum_{i=1}^{M} \delta n_i \epsilon_i \right) = 0.$

$$\left( \sum_{i=1}^{M} \delta n_i \ln n_i + \sum_{i=1}^{M} \delta n_i + \alpha \sum_{i=1}^{M} \delta n_i + \beta \sum_{i=1}^{M} \delta n_i \epsilon_i \right) = \sum_{i=1}^{M} (\alpha + \beta \epsilon_i + \ln n_i) \delta n_i = 0.$$

So, we get: $\alpha + \beta \epsilon_i + \ln n_i = 0$, or [54]:

(11) $$n_i = e^{-\alpha} e^{-\beta \epsilon_i}$$

---

[53] Using, instead, the more accurate (and the one more typically called Sterling's approximation) $\ln m! = m \ln m - m$ gives the exact same result as found in the ensuing calculations. This more accurate approximation estimates the factorial within 14% for $n_i = 10$.

[54] So, the probability of finding a particle in the $i^{th}$ cell is: $p(\epsilon_i) \equiv \frac{n_i}{N} = \frac{e^{-\alpha} e^{-\beta \epsilon_i}}{e^{-\alpha} \sum_{1}^{M} e^{-\epsilon_i/kT}} = \frac{e^{-\beta \epsilon_i}}{\sum_{1}^{M} e^{-\beta \epsilon_i}}$. This is

generalized to continuum case, as: $p(\epsilon) \equiv \frac{n(\epsilon)}{N} = \frac{e^{-\epsilon/kT} d\epsilon}{\int_{0}^{E_i} e^{-\epsilon/kT} d\epsilon} = \frac{e^{-\epsilon/kT} d\epsilon}{kT(1 - e^{-E_i/kT})}$.

(Note: if we took the more accurate version of Sterling's approximation, i.e. $\ln m! \approx m \ln m - m + \frac{1}{2}\ln 2\pi m$, we would get: $n_i e^{\frac{1}{4\pi n_i}} = e^{-\alpha} e^{-\beta \epsilon_i}$. This is actually $M$ equations, one for each of the $n_i$'s. We take $\epsilon_i \equiv i E_1/M$ and we also take $M$ to be fixed. $a \equiv e^{-\alpha}$ is a normalization factor that is fixed by the number constraint: $N = \sum_1^M n_i$. We also have the energy constraint:

$E_1 = \sum_1^M n_i \epsilon_i$. So, we have $M+2$ equations and $M+2$ unknowns ($M$ $n_i$'s, $a$, $\beta$ and $E_1$). In theory, we could solve the system of equations iteratively as follows. For a given, $n_i$, say $n_1$, we substituting chosen values for $M$ and $E_1$ and, we pick a trial $a$ and a trial $\beta$ and numerically find $n_i$ (say by graphical methods). We, then, do the same to find all the $n_i$'s. These should satisfy the number and energy constraints. If they do not we iterate our choices for $a$ and $\beta$ until the result comes out correctly.

The important point here is that for fixed $E_1$, we have a single $\beta$. That is, there is, even for low $N$, a one to one correspondence the total system energy and $\beta$ (note: $\beta$ characterizes the highest probability energy-binning state and will become $1/kT$ in the large $N$ limit.)

Now, we'd like to take the continuum limit of the bins, which involves $M \to \infty$. To make this easier, shift the sums above to sums from $i=0$ to $i=M-1$. This means, applying our system of binning (*which could have been anything up till now*) that we gave above, we get: $\epsilon_i = iE_1/M$. We thus have:

(12) $$n_i = e^{-\alpha} e^{-\beta i E_1/M}, \text{ so } N = e^{-\alpha} \sum_{i=0}^{M-1} e^{-\beta i E_1/M}.$$

where we take $a = e^{-\alpha}$. Then, noting that the geometric series partial sum is: $\sum_{i=0}^{M-1} ar^i = a\left(\frac{1-r^M}{1-r}\right)$, taking $r = e^{-\beta E_1/M}$, we get: $N = a\left(\frac{1-e^{-\beta E_1}}{1-e^{-\beta E_1/M}}\right)$; this implies: $a = N\left(\frac{1-e^{-\beta E_1/M}}{1-e^{-\beta E_1}}\right)$. Note how:

$$n_i = \lim_{M \to \infty} a e^{-\beta i E_1/M} \sim \lim_{M \to \infty} N\left(\frac{1-(1-\beta E_1/M)}{1-e^{-\beta E_1}}\right) e^{-\beta i E_1/M} = \lim_{M \to \infty} \frac{N}{M}\left(\frac{\beta E_1}{1-e^{-\beta E_1}}\right) \to 0.$$

As the bin gets arbitrarily small (i.e., $M \to \infty$), this is to be expected as this is the only way that the total number of particles can be finite as the number of bins approaches infinity. But, recalling our earlier requirement, we want $N \gg M$, so to maintain this (and thus also maintain our approximation that $n_i$ is large), we would have to let $N$ also go to infinity. We want to go to the continuum limit in the energy-binning, but, of course, we do not want to have an infinite number of particles. This leads us to back to the core of the matter; the bin occupation number $n_i$ is a way of asking how many particles are in a given energy range (with a fixed number of particles). This fact is implicit in the case of a finite number ($M$) of bins. We need to be more explicit in the limiting case.

Thus, to take the $M \to \infty$ limit, we consider, instead of the number of particles in a bin, the number of particles with a given energy in an arbitrarily small region $\Delta \epsilon$. That is, we define:

(13) $$\rho \equiv \lim_{\Delta \epsilon \to 0} \frac{n_i}{\Delta \epsilon} = \lim_{M \to \infty} \frac{a e^{-\beta i E_1 / M}}{E_1 / M}.$$

Now, calculating the total number of particles, we get, substituting for $a$ and using the Riemann definition of an integral:

(14) $$\text{Total number of particles} = \lim_{M \to \infty} \frac{M}{E_1} \sum_{i=0}^{M-1} a e^{-\beta i E_1 / M} \frac{E_1}{M}$$

$$= \lim_{M \to \infty} \frac{M}{E_1} \sum_{i=0}^{M} \frac{N}{M} \left( \frac{\beta E_1}{1 - e^{-\beta E_1}} \right) e^{-\beta (i E_1 / M)} \frac{E_1}{M}$$

$$= N \left( \frac{\beta}{1 - e^{-\beta E_1}} \right) \int_0^{E_1} e^{-\beta \epsilon} d\epsilon = N \left( \frac{\beta}{1 - e^{-\beta E_1}} \right) \left( \frac{1 - e^{-\beta E_1}}{\beta} \right) = N$$

which confirms our calculation by giving a result of $N$.

Now, by comparing $N = \int_0^{E_1} \rho(\epsilon) d\epsilon$ with the last line above, we can write the continuum limit of our binning result:

(15) $$\rho(\epsilon) = N \left( \frac{\beta}{1 - e^{-\beta E_1}} \right) e^{-\beta \epsilon}$$

Connecting this to the empirically verified Maxwell Boltzmann distribution validates our assumptions and gives us $\beta = 1/kT$, so that we have:

(16) $$\rho(\epsilon) = \frac{N}{kT} \frac{1}{\left(1 - e^{-E_1/kT}\right)} e^{-\epsilon/kT}$$

Note the total energy is:

(17) $$E_1 = \int_0^{E_1} \epsilon \rho(\epsilon) d\epsilon = \frac{N}{kT} \frac{1}{\left(1 - e^{-E_1/kT}\right)} \int_0^{E_1} \epsilon e^{-\epsilon/kT} d\epsilon = \frac{N E_1}{\left(1 - e^{E_1/kT}\right)} + NkT$$

For fixed $T$, graphing, $E_1$ and $\frac{N E_1}{\left(1 - e^{E_1/kT}\right)} + NkT$ reveals that there is a solution at $E_1 = 0$, which is physically invalid and there is a solution near[55]

(18) $$E_1 = NkT \text{ for large } N.$$

---

[55] Note, we are characterizing the whole system of $N$ particles, so we focus on $T$ and $E_1$; if we were interested in the individual particle states, we would also note that the average energy of an individual particle is: $\langle E \rangle = E_1 / N$.

At $E_1 = NkT$, $E_1 - \left( \frac{NE_1}{\left(1 - e^{E_1/kT}\right)} + NkT \right) = \frac{N^2 kT}{e^N - 1}$. So, for small N, this solution is far off but for our case of large N, it is a good approximation; in particular, $\frac{N^2 kT}{e^N - 1} << E_1$ implies, for $N >> 1$, $N^2 << \frac{e^N - 1}{kT} E_1 \approx e^N \left( \frac{E_1}{kT} \right)$   or   $N^2 e^{-N} << \left( \frac{E_1}{kT} \right)$. For macroscopic values, i.e. $N \approx 6.02 \times 10^{23} \approx 10^{24}$, the log of the left hand side is: $2\ln N - N \approx -10^{24}$, so $N^2 e^{-N} \approx 10^{-10^{24}}$, an incredibly small number. Hence, only when the temperature is *extremely* large or the total system energy is *extremely* small is this solution not approximately valid.

As mentioned in the text of the article, the important general point is that: *for an isolated idea gas (as described in text) at given a temperature, there is a single value for the total energy, establishing the correspondence between temperature and total energy.*

## Appendix A-2
(imposing a different binning requirement)

Let's see what happens if we let our system of total energy $E_1$ be such that no single particle can have an energy less than $\varepsilon_0$. We proceed as above.

To take the limit of $M \to \infty$, we first apply this different binning to get: $\epsilon_i = \varepsilon_0 + i(E_1 - \varepsilon_0)/M$. We thus have, using equation $n_i = e^{-\alpha} e^{-\beta \epsilon_i}$:

(19) $$n_i = e^{-\alpha} e^{-\beta(\varepsilon_0 + i(E_1 - \varepsilon_0)/M)} = e^{-(\alpha + \beta \varepsilon_0)} e^{-\beta i (E_1 - \varepsilon_0)/M},$$

so $N = a \sum_{i=0}^{M-1} e^{-i\beta(E_1 - \varepsilon_0)/M}$, where we take $a = e^{-(\alpha + \beta \varepsilon_0)}$.

Then, noting that the geometric series partial sum is: $\sum_{i=0}^{M-1} ar^i = a \left( \frac{1 - r^M}{1 - r} \right)$, taking $r = e^{-\beta(E_1 - \varepsilon_0)/M}$ and, we get: $N = a \left( \frac{1 - e^{-\beta(E_1 - \varepsilon_0)}}{1 - e^{-\beta(E_1 - \varepsilon_0)/M}} \right)$; this implies: $a = N \left( \frac{1 - e^{-\beta(E_1 - \varepsilon_0)/M}}{1 - e^{-\beta(E_1 - \varepsilon_0)}} \right)$.

Note how:

$$n_i = \lim_{M \to \infty} a e^{-\beta i E_1/M} \sim \lim_{M \to \infty} N \left( \frac{1 - (1 - \beta E_1/M)}{1 - e^{-\beta E_1}} \right) e^{-\beta i E_1/M} = \lim_{M \to \infty} \frac{N}{M} \left( \frac{\beta E_1}{1 - e^{-\beta E_1}} \right) e^{-\beta i E_1/M} \to 0$$

$$n_i = \lim_{M \to \infty} a e^{-\beta i (E_1 - \varepsilon_0)/M} \sim \lim_{M \to \infty} N \left( \frac{1 - (1 - \beta(E_1 - \varepsilon_0)/M)}{1 - e^{-\beta(E_1 - \varepsilon_0)}} \right) e^{-\beta i (E_1 - \varepsilon_0)/M}$$

$$= \lim_{M \to \infty} \frac{N}{M} \left( \frac{\beta(E_1 - \varepsilon_0)}{1 - e^{-\beta(E_1 - \varepsilon_0)}} \right) e^{-\beta i (E_1 - \varepsilon_0)/M} \to 0$$

As the bin gets arbitrarily small (i.e., $M \to \infty$), this is to be expected as this is the only way that the total number of particles can be finite as the number of bins approaches infinity. As above, to take the $M \to \infty$ limit, we thus consider instead of the number of particles in a bin, the number of particles with a given energy in an arbitrarily small region $\Delta\epsilon$. That is, we define, using $n_i = ae^{-\beta i(E_1-\varepsilon_0)/M}$:

(20) $$\rho \equiv \lim_{\Delta\epsilon \to 0} \frac{n_i}{\Delta\epsilon} = \lim_{M \to \infty} \frac{ae^{-\beta i(E_1-\varepsilon_0)/M}}{(E_1-\varepsilon_0)/M}.$$

Now, calculating the total number of particles, we get, substituting for $a$ and using the Riemann definition of an integral:

(21) Total number of particles $= \lim_{M \to \infty} \frac{M}{(E_1-\varepsilon_0)} \sum_{i=0}^{M-1} ae^{-\beta i(E_1-\varepsilon_0)/M} \frac{(E_1-\varepsilon_0)}{M}$

$$= \lim_{M \to \infty} \frac{M}{(E_1-\varepsilon_0)} \sum_{i=0}^{M-1} \frac{N}{M}\left(\frac{\beta(E_1-\varepsilon_0)}{1-e^{-\beta(E_1-\varepsilon_0)}}\right) e^{-\beta i(E_1-\varepsilon_0)/M} \frac{(E_1-\varepsilon_0)}{M}$$

$$= N\left(\frac{\beta}{1-e^{-\beta(E_1-\varepsilon_0)}}\right) \int_0^{E_1-\varepsilon_0} e^{-\beta\epsilon} d\epsilon = N\left(\frac{\beta}{1-e^{-\beta(E_1-\varepsilon_0)}}\right) \int_{\varepsilon_0}^{E_1} e^{-\beta(\epsilon-\varepsilon_0)} d\epsilon$$

$$= N\left(\frac{\beta}{1-e^{-\beta(E_1-\varepsilon_0)}}\right)\left(\frac{1-e^{-\beta(E_1-\varepsilon_0)}}{\beta}\right) = N$$

which confirms our calculation by giving a result of $N$.

Now, by comparing $N = \int_0^{E_1} \rho(\epsilon) d\epsilon$ with the last line above, we can write the continuum limit of our binning result, using $\beta = 1/kT$:

(22) $$\rho(\epsilon) = \frac{N}{kT}\left(\frac{1}{1-e^{-(E_1-\varepsilon_0)/kT}}\right) e^{-(\epsilon-\varepsilon_0)/kT}$$

Note the total energy is:
(23)
$$E_1 = \int_{\varepsilon_0}^{E_1} \epsilon \rho(\epsilon) d\epsilon = \frac{N}{kT}\left(\frac{1}{1-e^{-(E_1-\varepsilon_0)/kT}}\right) \int_{\varepsilon_0}^{E_1} \epsilon e^{-(\epsilon-\varepsilon_0)/kT} d\epsilon = \frac{N\left(E_1-\varepsilon_0 e^{(E_1-\varepsilon_0)/kT}\right)}{\left(1-e^{(E_1-\varepsilon_0)/kT}\right)} + NkT$$

For fixed $T$, graphing, $E_1$ and $\dfrac{N\left(E_1-\varepsilon_0 e^{(E_1-\varepsilon_0)/kT}\right)}{\left(1-e^{(E_1-\varepsilon_0)/kT}\right)} + NkT$ reveals that there is a solution at

$E_1 = \varepsilon_0$, which is physically invalid and there is, again, a solution near:
(24) $\qquad\qquad E_1 = NkT + \varepsilon_0$ for large $N$.

At $E_1 = NkT + \varepsilon_0$, $E_1 - \left(\dfrac{N\left(E_1-\varepsilon_0 e^{(E_1-\varepsilon_0)/kT}\right)}{\left(1-e^{(E_1-\varepsilon_0)/kT}\right)} + NkT\right) = \dfrac{N(N(\varepsilon_0+kT)-\varepsilon_0)}{e^{N(\varepsilon_0/kT+1)-\varepsilon_0/kT}-1}$. So, for

small $N$, this solution if far off but for our case of large N, it is a good approximation; in particular, $\dfrac{N(N(\varepsilon_0+kT)-\varepsilon_0)}{e^{N(\varepsilon_0/kT+1)-\varepsilon_0/kT}-1} << E_1$ implies, for

$N>>1$, $N^2 << E_1 \dfrac{e^{N(\varepsilon_0/kT+1)}-1}{(\varepsilon_0+kT)} \approx E_1 \dfrac{e^{N(\varepsilon_0/kT+1)}}{(\varepsilon_0+kT)}$ , or $N^2 e^{-N(\varepsilon_0/kT+1)} << \dfrac{E_1}{(\varepsilon_0+kT)}$,

Hence, for macroscopic size $N$, as before, only when the temperature is *extremely* large or the total system energy is *extremely* small is this solution not approximately valid. (Note: for small, i.e., $\varepsilon_0 << kT$, the inequality becomes: $N^2 e^{-N} << \dfrac{E_1}{kT}$, i.e., the same as before. For $\varepsilon_0 >> kT$, we have $N^2 e^{-N\varepsilon_0/kT} << \dfrac{E_1}{\varepsilon_0}$; the smallest the right hand side can become is one since $E_1 \geq \varepsilon_0$.)

Again, a key general point is that for a given temperature, there is always a single value of $E_1$ corresponding to it for any fixed $N$.